\documentclass[12pt]{article}
\usepackage{graphicx,amsmath,amssymb}
\usepackage{indentfirst}
\usepackage[usenames]{color}
\usepackage[colorlinks=true, urlcolor=navyblue, linkcolor=navyblue, citecolor=navyblue]{hyperref}
\usepackage{epstopdf}
\usepackage{enumerate}
\usepackage{appendix}
\usepackage{lineno}
\usepackage{setspace}

\definecolor{navyblue}{rgb}{0,0.08,0.45}

\begin{document}

\begin{flushright}
{\small JLAB-PHY-16-2312\\
SLAC-PUB-16772 \\ \vspace{2pt}}
\end{flushright}

\vspace{0pt}

\begin{center}
{\huge  Determination of  $\Lambda_{\overline{MS}}$ at Five}

\vspace{10pt}

{\huge   Loops from Holographic  QCD}

\end{center}

\vspace{10pt}

\centerline{Alexandre Deur\footnote{Contact author}}

\vspace{3pt}

\centerline{\it  Thomas Jefferson National Accelerator Facility, Newport
News, VA 23606, USA~\footnote{{\href{deurpam@jlab.org}{\tt
deurpam@jlab.org}}}}

\vspace{6pt}

\centerline{Stanley J. Brodsky}

\vspace{3pt}

\centerline {\it SLAC National Accelerator Laboratory, Stanford
University, Stanford, CA 94309,
USA~\footnote{{\href{mailto:sjbth@slac.stanford.edu}{\tt
sjbth@slac.stanford.edu}}}}

\vspace{6pt}

\centerline{Guy F. de T\'eramond}

\vspace{3pt}

\centerline {\it Universidad de Costa Rica, 11501 San Pedro de Montes de Oca, Costa
Rica~\footnote{{\href{mailto:gdt@asterix.crnet.cr}{\tt
gdt@asterix.crnet.cr}}}}

\vspace{10pt}

\begin{abstract}
The recent determination of the $\beta$--function of the QCD running coupling $\alpha_{\overline{MS}}(Q^2)$ to 5-loops{\color{blue},} provides a verification of the convergence of a novel method for determining the fundamental QCD parameter $\Lambda_s$ based on the Light-Front Holographic approach to nonperturbative QCD. 
The new 5-loop analysis, together with improvements in determining the holographic QCD nonperturbative scale parameter $\kappa$ from hadronic spectroscopy, leads to an improved precision of the value of $\Lambda_s$ in the ${\overline{MS}}$ scheme close to a factor of two; we find $\Lambda^{(3)}_{\overline{MS}}=0.339\pm0.019$ GeV for $n_{f}=3$, in excellent agreement with the world average, $\Lambda_{\overline{MS}}^{(3)}=0.332\pm0.017$ GeV. We also discuss the constraints imposed on the scale dependence of the strong coupling in the nonperturbative  domain by superconformal quantum mechanics and its holographic embedding in anti-de Sitter space.
\end{abstract}

The strong coupling $\alpha_{s}$ is a central quantity for the study of Quantum Chromodynamics (QCD), the gauge theory of the strong interactions~\cite{Deur:2016tte}. Traditionally,
 $\alpha_s$ --or equivalently, the perturbative QCD (pQCD) scale parameter $\Lambda_s$--   has  been determined from measurements of high momentum processes or from Lattice Gauge Theory.  More recently,  $\Lambda_s$ has also been determined from nonperturbative dynamics using  light-front holographic QCD  (LFHQCD)~\cite{Deur:2014qfa}, an approach to color confinement that successfully describes both the hadronic spectrum and the bound-state light-front wave functions that control hadronic processes~\cite{Brodsky:2014yha}.

 This new approach to hadron physics is based on superconformal quantum mechanics~\cite{Fubini:1984hf,deTeramond:2014asa,Dosch:2015nwa} and its light-front (LF) holographic embedding in a higher dimensional gravitational theory~\cite{Brodsky:2006uqa,deTeramond:2008ht,deTeramond:2013it}. The result is a semiclassical effective theory which incorporates fundamental aspects of nonperturbative QCD that are not apparent from its classical Lagrangian, such as the emergence of a mass scale and confinement~\cite{Brodsky:2013ar}, the existence of a zero-mass bound state~\cite{Dosch:2015nwa},  the appearance of universal Regge trajectories and the breaking of chiral symmetry~\cite{Dosch:2015nwa,deTeramond:2016htp}. In addition, it gives remarkable connections between the light meson and nucleon spectra~\cite{Dosch:2015nwa}. Only one mass parameter appears  -- the confinement scale $\kappa$, which is constrained to better than 5\% by measurements of hadron masses and other hadronic observables~\cite{Brodsky:2016yod}. 

It has been recognized  before the advent of QCD, that the linearity of the Regge trajectories implies oscillatory modes 
of constituent quarks within the hadron~\cite{Minkowski:1971wp,Feynman:1971wr}. The subsequent exploration of 
covariant two-particle Hamiltonians in the null plane lead uniquely to relativistic harmonic confinement if the wave 
equations are local differential equations~\cite{Leutwyler:1977vy}. These general results were extended to the case of 
spin-$\frac{1}{2}$ constituents in Ref.~\cite{Leutwyler:1977pv}. 

As shown  in a remarkable article by de Alfaro, Fubini and Furlan 
(dAFF)~\cite{deAlfaro:1976vlx},  it is possible to generate a mass 
scale $\kappa$ and a confinement potential  while maintaining the conformal  symmetry of the 
action.  In~\cite{deAlfaro:1976vlx} dAFF write the quantum mechanical evolution operator  as a superposition of 
the generators of the conformal group $Conf\left(R^1\right)$: The generator of 
time translation $H$, the generator of dilatations $D$, and the generator of special 
conformal transformations $K$. Since the generators of  $Conf\left(R^1\right)$ 
have different dimensions, a mass scale is introduced which in the present 
context plays a fundamental role, as initially conjectured in  Ref.~\cite{deAlfaro:1976vlx}.
The dAFF mechanism was extended to superconformal quantum mechanics in  Refs.~\cite{Fubini:1984hf,Akulov:1984uh}.
One can reduce the LF Hamiltonian equations in QCD for massless quarks to a single-variable LF 
Schrodinger equation in $\zeta$, identical to the equations derived from AdS$_5$ in the variable $z$.
The color confining potential is unique using the dAFF procedure. It has the form of a harmonic oscillator $\kappa^4 \zeta^2$.
In LFHQCD, the soft-wall dilaton, which encodes the breaking of conformal symmetry 
in the higher dimensional anti-de Sitter  AdS$_5$ space must thus have the 
form $e^{\kappa^2 z^2}$. The  holographic variable  $z$  in the 5-dimensional classical gravity theory 
is identified with the invariant transverse separation $\zeta$ between the hadron
constituents in the light-front quantization scheme~\cite{Brodsky:2006uqa,Dirac:1949cp,Brodsky:1997de}.
The harmonic form of the confining light-front potential is equivalent to  the familiar linear 
heavy quark $Q\bar Q$ potential in the instant form~\cite{Trawinski:2014msa}  and has been 
successful in reproducing essential nonperturbative QCD features, such
as Regge  trajectories  and the $Q^2$-dependence of hadronic form factors~\cite{Brodsky:2014yha}.

In Quantum Field Theory, couplings acquire a scale-dependence due to short-distance quantum 
effects   which are included in their definition. In particular for $\alpha_s$, the running is  determined 
by pQCD and its renormalization group equation~\cite{Gross:1973id}. Likewise, the scale dependence 
of  $\alpha_s$ in the  nonperturbative domain can be obtained from the large-distance confining 
potential and, in the LFHQCD framework, follows from the specific embedding of  light-front dynamics 
in anti-de Sitter (AdS) space~\cite{Brodsky:2010ur}. Its specific form is obtained from the dilaton profile 
which breaks conformal invariance in the AdS$_5$ action: It is uniquely determined from the constraints 
imposed by the superconformal algebraic structure~\cite{deTeramond:2014asa,Dosch:2015nwa,deTeramond:2016htp}.  
The matching of the short- and large-distance regimes of the strong coupling  $\alpha_s$ determines 
the QCD perturbative scale $\Lambda_s$ in any renormalization scheme in terms of the physical hadronic 
scale $\kappa$~\cite{Deur:2016cxb}. The procedure also sets the scale separating perturbative and 
nonperturbative hadron dynamics.
We remark that since $\kappa$ is a physical parameter, it cannot depend on the choice of renormalization 
scheme, contrary to $\Lambda_s$. In fact, perturbative renormalization or evolution is not relevant to 
LFHQCD and $\kappa$. It is worth mentioning that some nonperturbative approaches, such as Lattice 
Gauge Theory, do become scheme-dependent because they are matched to perturbative results in order 
to fix parameters, but this is not the case of LFHQCD where $\kappa$ is fixed by observables, i.e.,
scheme-independent quantities. In our procedure which uses a scheme-independent nonperturbative 
formalism, the scheme-dependence of $\Lambda_s$ emerges from the infrared fixed-point value of 
$\alpha_s$, which is RS-dependent (the running coupling is not an observable, and it is thus scheme-dependent) 
and is not predicted by LFHQCD. LFHQCD predicts only the scale-dependence of $\alpha_s$. The 
infrared fixed-point value is determined in a particular scheme (the $g_1$ scheme) using a sum rule~\cite{Brodsky:2010ur}. 
The values in other schemes, e.g., $\overline{MS}$, are then obtained using 
``Commensurate Scale Relations"~\cite{Brodsky:1994eh}, which are strict predictions of pQCD.

The method used to derive  $\Lambda_s$ from  LFHQCD uses the  effective charge $\alpha_{g_{1}}$, defined from the Bjorken sum rule~\cite{Bjorken:1969mm}.
It has the  analytic form~\cite{Brodsky:2010ur}:
\begin{equation} \label{alphaIR}
{ \frac{\alpha^{IR}_{g_{1}}\left(Q^{2}\right)}{\pi} }  = \exp \left( - {\frac{Q^2} {4\kappa^2}} \right),
\end{equation}
in the infrared (IR) nonperturbative regime{\color{blue}.}
Here $Q$ is the momentum transfer in the spin-dependent nucleon structure functions appearing in the Bjorken sum rule, and
$\kappa$ is the fundamental LFHQCD scale parameter determined from the light hadron spectrum.
This prediction for $\alpha^ {IR}_{g_{1}}(Q^{2})$ agrees remarkably well 
with experimental data for $\alpha_{g_{1}}(Q^{2})$ in the domain
$Q^2  \leq 1~$GeV$^2$~\cite{Deur:2005cf} where  LFHQCD is applicable, and it displays an infrared fixed point. 
In the nonperturbative domain, the relations between $\alpha_{g_{1}}(Q^{2})$ 
and  the strong couplings $\alpha_s(Q^2)$ in other  renormalization schemes, such 
as the $\overline{MS}$, MOM, or V schemes are given in Ref.~\cite{Deur:2016cxb}. Such relations are obtained
by first assuming that  $\alpha_s$ always has an infrared fixed point regardless of the scheme it is expressed in. Then,
$\alpha_s(Q^2=0)$ is left as a free parameter to be determined by the matching procedure 
described below, but with  the perturbative scale $\Lambda_s$ determined by the world data. 

The effective charge $\alpha_{g_{1}}$ can be expressed at high momentum transfer as a perturbative expansion 
in the perturbative coupling $\alpha_{\overline{MS}}(Q^{2})$, as 
defined by the  $\overline{MS}$ renormalization scheme~\cite{Deur:2005cf}:
\begin{equation}
\label{eq:msbar to g_1}
\alpha_{g_{1}}(Q^2)= \pi \left[ \frac{\alpha_{\overline{MS}}(Q^2)}{\pi} + a_1 \left(\frac{\alpha_{\overline{MS}}(Q^2)}{\pi}\right)^2 + 
a_2 \left(\frac{\alpha_{\overline{MS}}(Q^2)}{\pi}\right)^3  \cdots \right],
\end{equation}
with the coefficients  $a_i$ known up to $a_4$~\cite{Baikov:2010je} and $a_5$ having been only
estimated~\cite{Kataev:2005hv}.
The normalization and evolution of $\alpha_{g_{1}}$ is then determined in the $\overline{MS}$ 
renormalization scheme by the QCD $\beta_{\overline{MS}}$-function and the  
mass scale $\Lambda_{\overline{MS}}$~\cite{Gross:1973id}. 
Global hadron-parton duality~\cite{Bloom:1970xb} predicts that
the nonperturbative description for $\alpha_{g_{1}}(Q^{2})$ 
overlaps  with the pQCD expression at  intermediate 
values of $Q^2$. Matching the  LFHQCD and pQCD expressions of $\alpha_{g_{1}}(Q^{2})$
and their derivatives then allows us to determine $\Lambda_{\overline{MS}}$
and the scale $Q_{0}$ characterizing the transition between the perturbative and  nonpeturbative
descriptions. The comparison between $\Lambda_{\overline{MS}}$
obtained from  light-front holographic QCD and the world data  provides a key 
test of this novel approach to  nonperturbative QCD.

It is usually argued that one determines the proton mass and other aspects  of the QCD 
mass scale starting from a measurement of $\Lambda_s$ in  the pQCD domain. 
This ansatz is difficult to justify since $\Lambda_s$ is renormalization scheme 
dependent, whereas masses or other   physical  observables are not.  
In fact, the procedure outlined above is the  opposite: 
$\Lambda_s$ is determined  in any scheme 
starting from the fundamental --scheme independent-- confinement scale $\kappa$ of nonperturbative QCD.  
Since the QCD Lagrangian has no mass parameter  in the limit where the quark masses are neglected,
the magnitude of the mass parameter $\kappa$ cannot be determined in fixed units  by QCD itself.  
Actually, the units normally used for mass, GeV, are a convention. The key predictions are thus  
ratios such as   $\Lambda_s  / \kappa$.  
The value of $\kappa$ determines all other mass scales in the chiral limit.  
Indeed,  holographic QCD predicts the ratios of masses and mass times radius, etc.  For example, it predicts 
$m_p / \Lambda_s $~\cite{Deur:2014qfa}, $m_{\rho} /  m_p$, $m_p \times R_p$~\cite{Brodsky:2014yha}, etc. 
Thus  $\kappa$  is in a sense  a ``holding parameter",  a scale which arises from color confinement 
and the breaking of conformal symmetry, but it cannot be determined in absolute units by QCD.  
In fact, while the emergence of the QCD mass scale is attributed in LFHQCD 
to the dAFF symmetry breaking procedure, its value is essentially unknown, since 
the vacuum state in the dAFF construction is chosen {\it ab initio}.  A specific 
value for $\kappa$ is not determined by QCD alone. The scale only becomes fixed when we make a measurement 
such as the pion decay constant or the $\rho$ mass.  Thus QCD with massless quarks can only predict ratios such 
as $m_p / m_\rho = \sqrt{2}$. The dAFF mechanism also differs from spontaneous symmetry breaking 
or explicit symmetry breaking by adding mass terms to the Lagrangian.

Since our initial LFHQCD determination of $\Lambda_{\overline{MS}}$ reported in 
Ref.~\cite{Deur:2014qfa}, several new developments have occurred which allow us 
to efficiently test the convergence of our determination as well as significantly improve 
the comparison between light-front holographic QCD and the world data:
1)  The LFHQCD scale parameter $\kappa$ has been determined
with greater accuracy  from a systematic analysis of the light-quark excitation 
spectra~\cite{Brodsky:2016yod} in  the context of  the semiclassical 
superconformal approach unifying mesons and baryons~\cite{Dosch:2015nwa};
2)  The running of $\alpha_{\overline{MS}}(Q^{2})$ has been computed  to
five loops~\cite{Baikov:2016tgj}, that is the $\beta$--function is now known up
to order $\beta_4$  in the $\overline{MS}$ renormalization scheme; and 3) the average world data
for $\Lambda_{\overline{MS}}$ has been updated~\cite{Olive:2016xmw}.

In this  article, we improve our determination of $\Lambda_{\overline{MS}}$
from the  light-front holographic QCD framework~\cite{Deur:2014qfa} utilizing these 
new developments. We also study the convergence of this determination. The
pQCD approximants  are asymptotic Poincar\'{e}
series that converge up to an optimal order $\sim1/a$, where $a=\alpha^{pQCD}_{s}/\pi$
is the expansion parameter of the series.   Indeed, we have shown in Ref.~\cite{Deur:2016cxb} that the transition between the
 LFHQCD description of $\alpha_{s}(Q^{2})$ and its pQCD description
occurs at $Q_{0}^{2}=0.75\pm0.07$ GeV$^2$ in the $\overline{MS}$ scheme:
The optimal order in the Poincar\'e series is thus $1/a(Q_{0}^{2}) \simeq 8$.
Consequently, it is advantageous to use $\alpha_{\overline{MS}}^{pQCD}(Q^{2}>Q_{0}^{2})$
evaluated at five loops to obtain an accurate value of $\Lambda_{\overline {MS}}$ 
following the matching procedure with the nonperturbative regime described above.

\section{Result for $\Lambda_{\overline{MS}}$}

The perturbative series of the $\beta$ function
\begin{equation} 
Q^{2}\frac{\partial}{\partial Q^{2}} \frac{\alpha_{s}}{4\pi}=\beta\left(\alpha_{s}\right) = 
-\left(\frac{\alpha_{s}}{4\pi}\right)^{2}\sum_{n=0}\left(\frac{\alpha_{s}}{4\pi}\right)^{n}\beta_{n},
\end{equation}
calculated up to order $\beta_{4}$  yields  the five-loop expression of $\alpha_{\overline{MS}}^{pQCD}$~\cite{Kniehl:2006bg}:
\begin{multline} 
\alpha^{pQCD}_{\overline{MS}}(Q^2) = \frac{4\pi}{\beta_{0}t} \biggl[ 1-\frac{\beta_{1}}{\beta_{0}^{2}}\frac{\mbox{ln}(t)}{t}  
+ \frac{\beta_{1}^{2}}{\beta_{0}^{4}t^2}\left(\mbox{ln}^2(t)-\mbox{ln}(t)-1+\frac{\beta_{2}\beta_{0}}{\beta_{1}^{2}}\right) \\
+ \frac{\beta_{1}^{3}}{\beta_{0}^{6}t^{3}}\biggl(-\mbox{ln}^3(t)+\frac{5}{2}\mbox{ln}^2(t) 
+2 \, \mbox{ln}(t)-\frac{1}{2}-3\frac{\beta_{2}\beta_{0}}{\beta_{1}^{2}}\mbox{ln}(t)+\frac{\beta_{3}\beta_{0}^{2}}{2\beta_{1}^{3}}\biggr)\\
+ \frac {\beta_1^4} {\beta_0^8 t^4} \bigg(\mbox{ln}^4(t) -\frac{13} {3} \mbox{ln}^3(t) -\frac{3} {2}  \mbox{ln}^2(t)  +  
  4 \, \mbox{ln}(t) +\frac{7}{6} +  \frac {3\beta_2 \beta_0}{\beta_1^2} \left( 2 \, \mbox{ln}^2(t) - \mbox{ln}(t) -1\right) \\
  -  \frac{\beta_3\beta_0^2}{\beta_1^3}\left(2 \, \mbox{ln}(t) + \frac{1}{6}\right) +\frac{5\beta_2^2 \beta_0^2}{3\beta_1^4}+\frac{\beta_4 \beta_0^3}{3 \beta_0^4} \biggr)
+\mathcal{O}\left(\frac{\mbox{ln}(t)^{6}}{t}\right) \biggr] 
\label{eq:alpha_s},
\end{multline}
with $t=\mbox{ln}\left(Q^{2}/\Lambda_s ^{2}\right)$ and 
\begin{equation}
\beta_{0}=11-\frac{2}{3}n_{f},
\end{equation}
\begin{equation}
\beta_{1}=102-\frac{38}{3}n_{f},
\end{equation} 
\begin{equation}
\beta_{2} = \frac{2857}{2}-\frac{5033}{18}n_{f}+\frac{325}{54}n_{f}^{2},
\end{equation}
\begin{multline}
\beta_{3} = \left(\frac{149753}{6}+3564 \, \xi\left(3\right)\right)-\left(\frac{1078361}{162}+
\frac{6508}{27}\xi\left(3\right)\right)n_{f}+\left(\frac{50065}{162}+\frac{6472}{81} \, \xi
\left(3\right)\right)n_{f}^{2}   \\
+\frac{1093}{729}n_{f}^{3},
\end{multline}
and
\begin{multline}
\beta_4 = \frac{8157455}{16} + \frac{621885}{2}\xi_3 - \frac{88209}{2}\xi_4 -288090 \xi_5 + \\
 \bigg(-\frac{336460813}{1944} -\frac{4811164}{81}\xi_3 + \frac{33935}{6}\xi_4 + \frac{1358995}{27}\xi_5 \bigg) n_f + \\
 \bigg(\frac{25960913}{1944} + \frac{698531}{81}\xi_3 - \frac{10526}{9}\xi_4 - \frac{381760}{81}\xi_5 \bigg)n_f^2+ \\
 \bigg(-\frac{630559}{5832} -\frac{48722}{243}\xi_3 + \frac{1618}{27}\xi_4 + \frac{460}{9}\xi_5 \bigg)n_f^3+ \\
 \bigg(\frac{1205}{2916} -\frac{152}{81}\xi_3  \bigg)n_f^4,
\end{multline}
with $\xi_n$ the Riemann zeta function~\cite{Baikov:2016tgj}. The coefficients $\beta_0$ and $\beta_1$ are scheme independent and the higher order coefficients are given in the $\overline{MS}$  renormalization scheme.
Here, we will set $n_{f}=3$ and use the updated value of
the holographic QCD scale parameter, $\kappa=0.523\pm0.024$ GeV determined from the excitation spectra of all light mesons and baryons~\cite{Brodsky:2016yod}.
This value characterizes the mass scale of light-quark hadron spectroscopy 
and is compatible with the fit to the Bjorken sum data at low
$Q^{2}$~\cite{Deur:2004ti} in the holographic QCD validity domain, which
yields $\kappa=0.496\pm0.007$ GeV~\cite{Deur:2016tte}. 
The updated value of $\kappa$  is lower than  --but compatible with-- the 
value we used in~\cite{Deur:2014qfa}: $\kappa=m_{\rho}/ \sqrt 2 = 0.548$ GeV~\cite{Brodsky:2014yha}, with 
$m_{\rho}$ the $\rho$--meson mass. This value is also used in the study of  hadronic form factors, which are expressed in terms of $\rho$ mass poles and its radial recurrencies~\cite{Brodsky:2014yha,Sufian:2016hwn}.

As in Ref.~\cite{Deur:2014qfa}, we compute $\alpha_{g1}^{pQCD}(Q^{2})$
using the Bjorken sum rule~\cite{Bjorken:1969mm} up to 5th order in $\alpha_{\overline{MS}}^{pQCD}$
\cite{Baikov:2010je}. At $\beta_{4}$ and
$\left(\alpha_{\overline{MS}}^{pQCD}\right)^{4}$ orders, we obtain $\Lambda_{\overline{MS}}=0.339 \pm 0.019$
GeV and $Q_0^2 = 1.14 \pm 0.12$ GeV$^2$  by matching the nonperturbative and perturbative expressions for the couplings, Eqs. (\ref{alphaIR}) and (\ref{eq:msbar to g_1}) respectively. This value of $\Lambda_{\overline{MS}}$ is to be compared to the 
present world data, $\Lambda_{\overline{MS}}^{PDG}=0.332 \pm 0.017$
GeV for $n_{f}=3$~\cite{Olive:2016xmw}. 
(The value of $Q_0$ is given in the $g_1$ scheme and is higher than 
the corresponding value in the $\overline{MS}$ scheme~\cite{Deur:2016cxb}.)

The  uncertainties entering our determination
stem from the uncertainty on $\kappa$ ($\pm \,0.016$ GeV), the uncertainty from the chiral
limit approximation ($\pm \, 0.003$ GeV) and  the truncation
uncertainty on the Bjorken and  $\alpha_{\overline{MS}}^{pQCD}$ series, Eqs. (\ref{eq:msbar to g_1})
and (\ref{eq:alpha_s}), respectively ($\pm \, 0.010$ GeV).
This uncertainty is taken, for order $n$, as the difference between the results at orders $n$ and $n+1$, the uncertainty
at the highest order being taken equal to that of the preceding order.
The first two contributions to the total uncertainty reflect the consequence of approximations necessary to make
the LFHQCD approach tractable. One  could systematically 
improve LFHQCD towards exact QCD by diagonalizing  the true QCD LF Hamlitonian on an orthonormal basis constructed
from the AdS/QCD solutions. This is a method called BLFQ (Basis Light-Front Quantization)~\cite{Vary:2013kma}.

\begin{figure}[ht]
\centerline{\includegraphics[width=.60\textwidth]{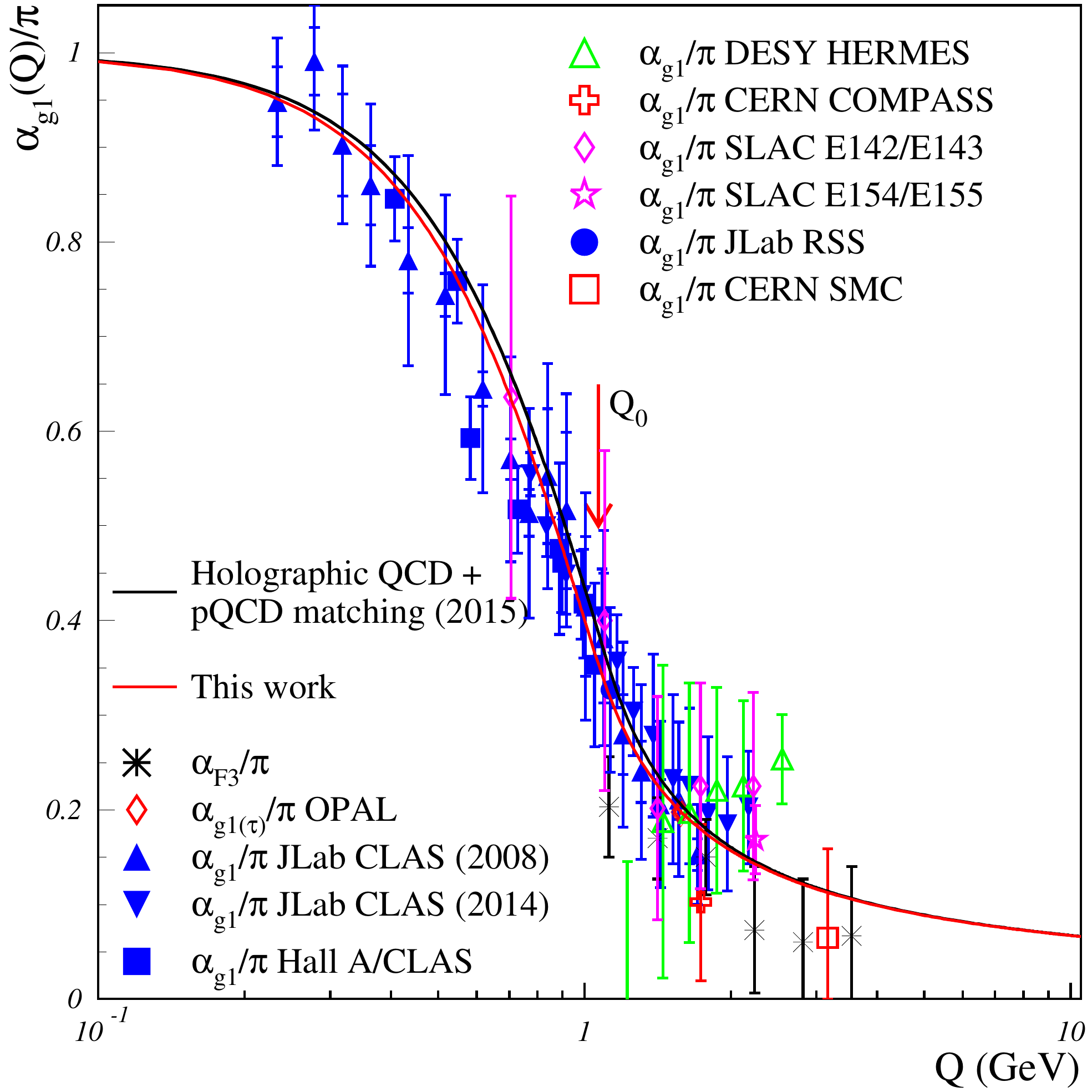}}
\caption{\label{Flo:comparison} Running of $\alpha_{g_1}(Q)$ for 
$\kappa=0.523$ GeV, $\Lambda_{\overline{MS}}=0.339$ GeV (red line). 
Also shown are experimental data~\cite{Deur:2005cf} and the 
earlier determination of $\alpha_{g_1}(Q)$ for $\kappa=0.548$ GeV, 
$\Lambda_{\overline{MS}}=0.341$ GeV (black line)~\cite{Deur:2014qfa}. 
The arrow marks the transition scale $Q_0$  to the perturbative regime.}
\end{figure}

The total uncertainty has significantly
improved compared to our previous determination, $\Lambda_{\overline{MS}}=0.341\pm0.032$
GeV~\cite{Deur:2014qfa}. The updated prediction 
of the running coupling is shown in Fig.~\ref{Flo:comparison},
together with the previous determination~\cite{Deur:2014qfa} and experimental data~\cite{Deur:2005cf}.
%
%

The result using the Bjorken sum rule coefficient $a_5$ 
in Eq. (\ref{eq:msbar to g_1}), which is assessed in Ref.~\cite{Kataev:2005hv}, is $\Lambda_{\overline{MS}}=0.317\pm0.019$
GeV. The uncertainty stems from the uncertainty on $\kappa$ ($\pm \, 0.015$ GeV),
the uncertainty from the chiral limit approximation ($\pm \, 0.003$ GeV), the truncation
uncertainty on the Bjorken and  $\alpha_{\overline{MS}}^{pQCD}$ series, Eqs. (\ref{eq:msbar to g_1})
and (\ref{eq:alpha_s}), respectively, ($\pm \, 0.010$ GeV), and an estimate on  the $a_5$ uncertainty ($\pm \, 0.005$ GeV). 
This latest contribution is assessed by rescaling $a_5$ by  the factor 175.7/130 and obtaining
$\Lambda_{\overline{MS}}$ with this rescaled value.  Indeed, the estimate of the Bjorken sum rule
coefficient $a_4$ in Ref.~\cite{Kataev:2005hv} was 130 while the recent exact calculation yields
$a_4=175.7$~\cite{Baikov:2010je}. The ratio 175.7/130  thus provides an indication of the uncertainty on $a_5$.
We will not quote $\Lambda_{\overline {MS}}$ at fifth order, since the coefficient $a_5$ in Eq.~\ref{eq:msbar to g_1} has only been estimated rather than computed.

\begin{figure}[ht]
\centerline{\includegraphics[width=.60\textwidth]{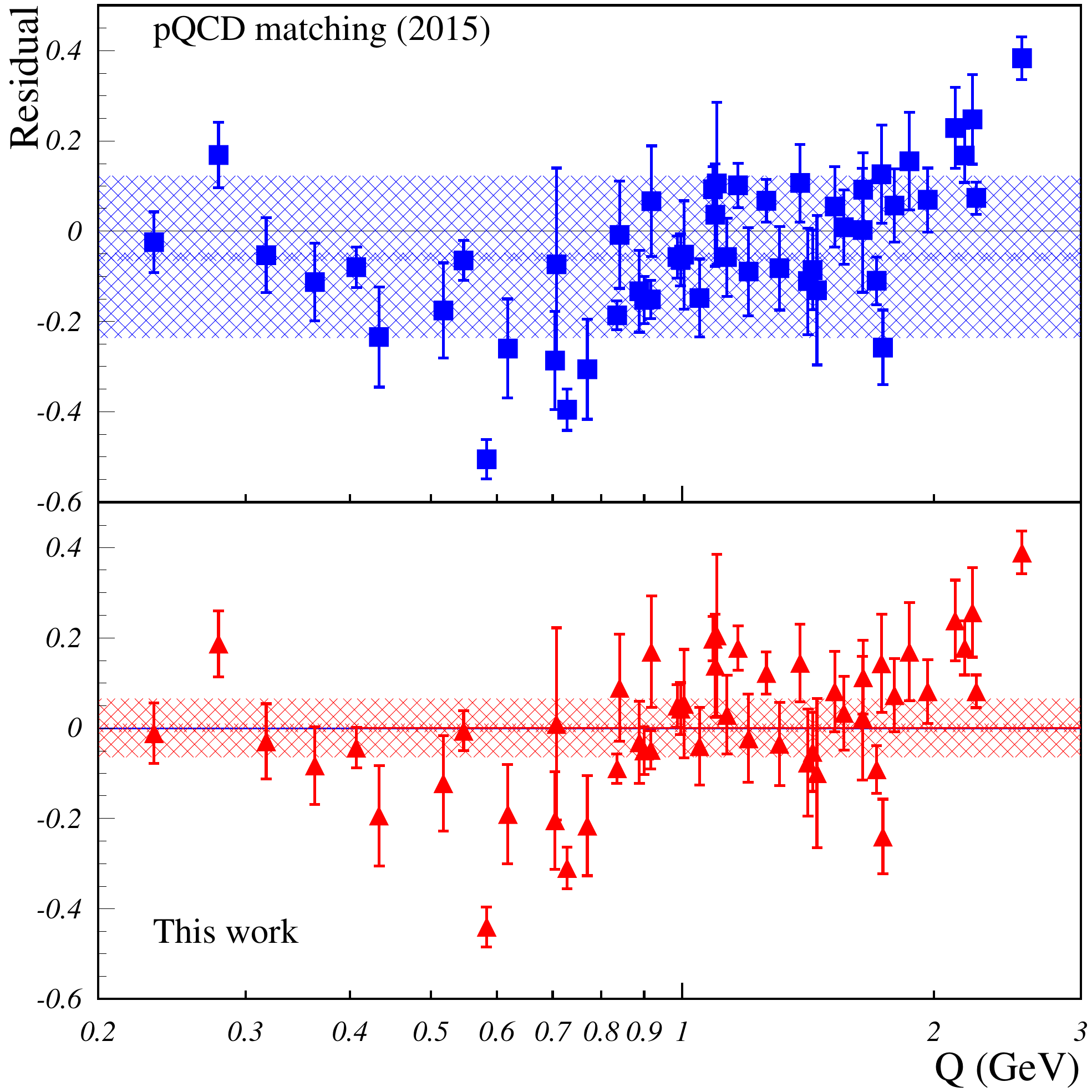}}
\caption{\label{Flo:residuals} Residual between the LFHQCD and pQCD matched  
$\alpha_{g_1}(Q)$ and the experimental data. The top panel  corresponds to the previous 
determination obtained in Ref.~\cite{Deur:2014qfa}. The bottom  pannel is our present result. 
In each panel, the broader band is the total uncertainty while the thiner and denser 
band inside represents the fit uncertainty only.}
\end{figure}

The present uncertainty on $\Lambda_{\overline{MS}}$ has improved 
by close to a factor of 2 compared to the result reported in Ref.~\cite{Deur:2014qfa}.
Another way to quantify the improvement between our previous determination 
and the present result is to inspect the residual between $\alpha_{g_1}(Q)$ obtained on the full $Q$-range using our
matching procedure and the experimental data.  We show  such residuals in Fig.~\ref{Flo:residuals}.
The matching procedure does not involve any fit to the experimental data and has no free parameter: $\kappa$ is fixed and $\Lambda_{\overline{MS}}$ is obtain from the 
matching, without influence from data. Thus, the departure from zero 
of the residual and the  $\chi^2$ of its averaged value quantify the
agreement between two determinations of $\alpha_{g_1}$ --from experiments, and from  the matching procedure of LFHQCD and pQCD described here-- 
that are fully independent. The averaged residual for the present result is 
$5.7\times10^{-4} \pm \, 9.2 \times10^{-3}$ (exp.) $\pm \, 6.4 \times10^{-2}$ (theo.) with $\chi^2 = 7.2$.
The result from Ref.~\cite{Deur:2014qfa} yields an averaged residual of 
$5.7\times10^{-2} \pm \, 9.2 \times 10^{-3}$ (exp.)\,$\pm\, 1.8 \times10^{-1}$ (theo.) with $\chi^2 = 8.3$.
The ``experimental" uncertainty reflects for the gaussian deviation of the experimental data from the average value of the residual.
It is evidently the same in both cases, since they use the same experimental data.
The theoretical  uncertainty reflects the uncertainty of the theoretical  prediction for $\alpha_{g_1}(Q)$ 
obtained with our matching procedure, that is the uncertainty on $\kappa$ and $\Lambda_{\overline{MS}}$.  
The size of the theoretical uncertainty improved significantly, by a factor of 3. It largely dominates the ``experimental" 
uncertainty. In addition, the residual value for the present result is much closer to zero compared to the   result obtained in
Ref.~\cite{Deur:2014qfa}, although both of them are compatible with zero. The $\chi^2$  is also improved, albeit marginally. 
In all, these comparisons quantify the significant improvement of our determination of $\Lambda_{\overline{MS}}$ and
 consequently  of $\alpha_{g_1}(Q)$ compared to our earlier result.

\section{Convergence}

\begin{figure}[ht]
\centerline{\includegraphics[width=.60\textwidth]{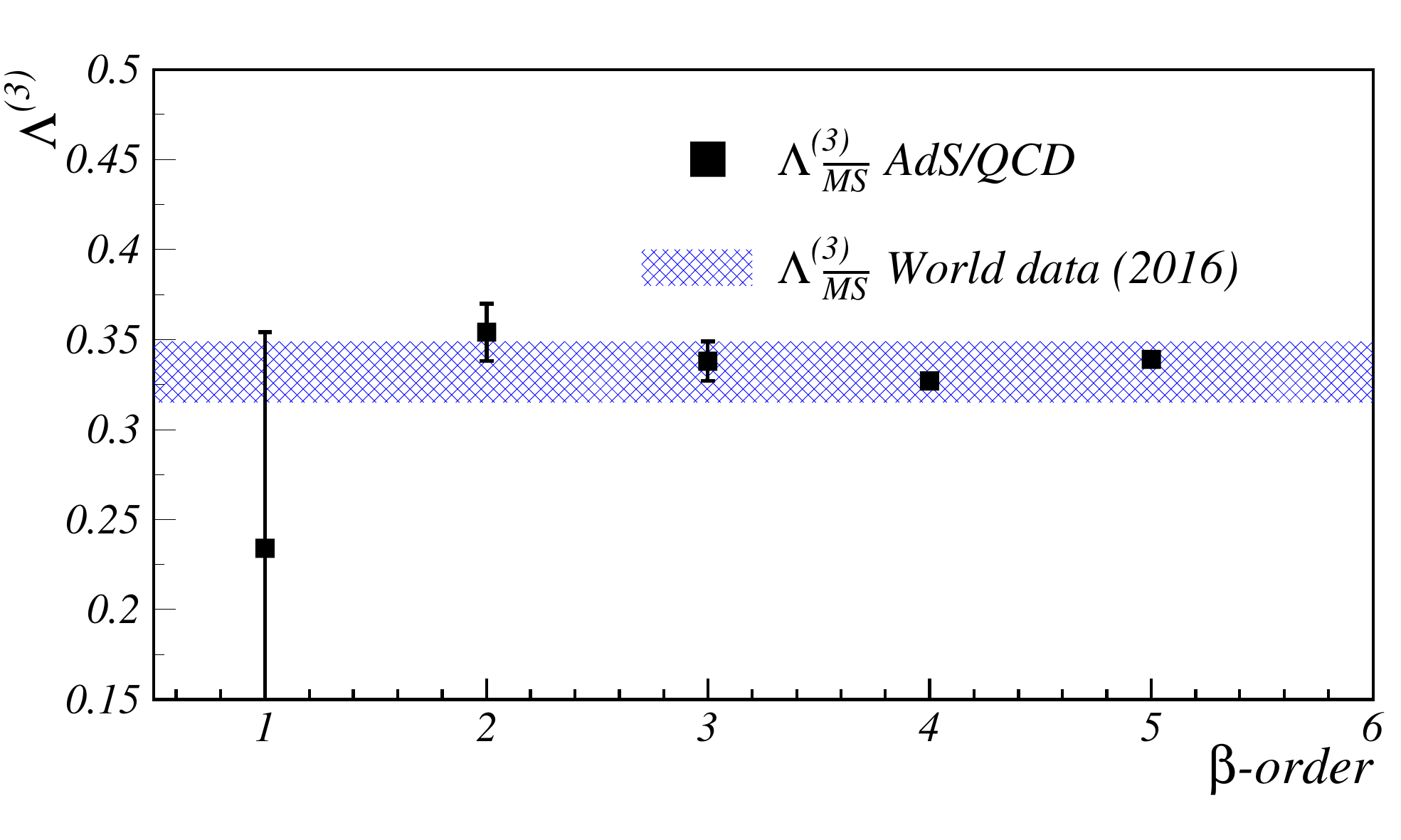}}
\caption{\label{Flo:convergence_beta} Convergence of our determination of $\Lambda_{\overline{MS}}$
(black squares)  as a function of the $\beta$-series order for $n_{f}=3$.
The pQCD series for the Bjorken sum rule  is computed at order $\left(\alpha_{\overline{MS}}^{pQCD}\right)^{4}$.
The error bars reflect only  the uncertainty from the truncation of the $\beta$-series. The blue band gives the latest world data.}
\end{figure}

\begin{figure}[ht]
\centerline{\includegraphics[width=.60\textwidth]{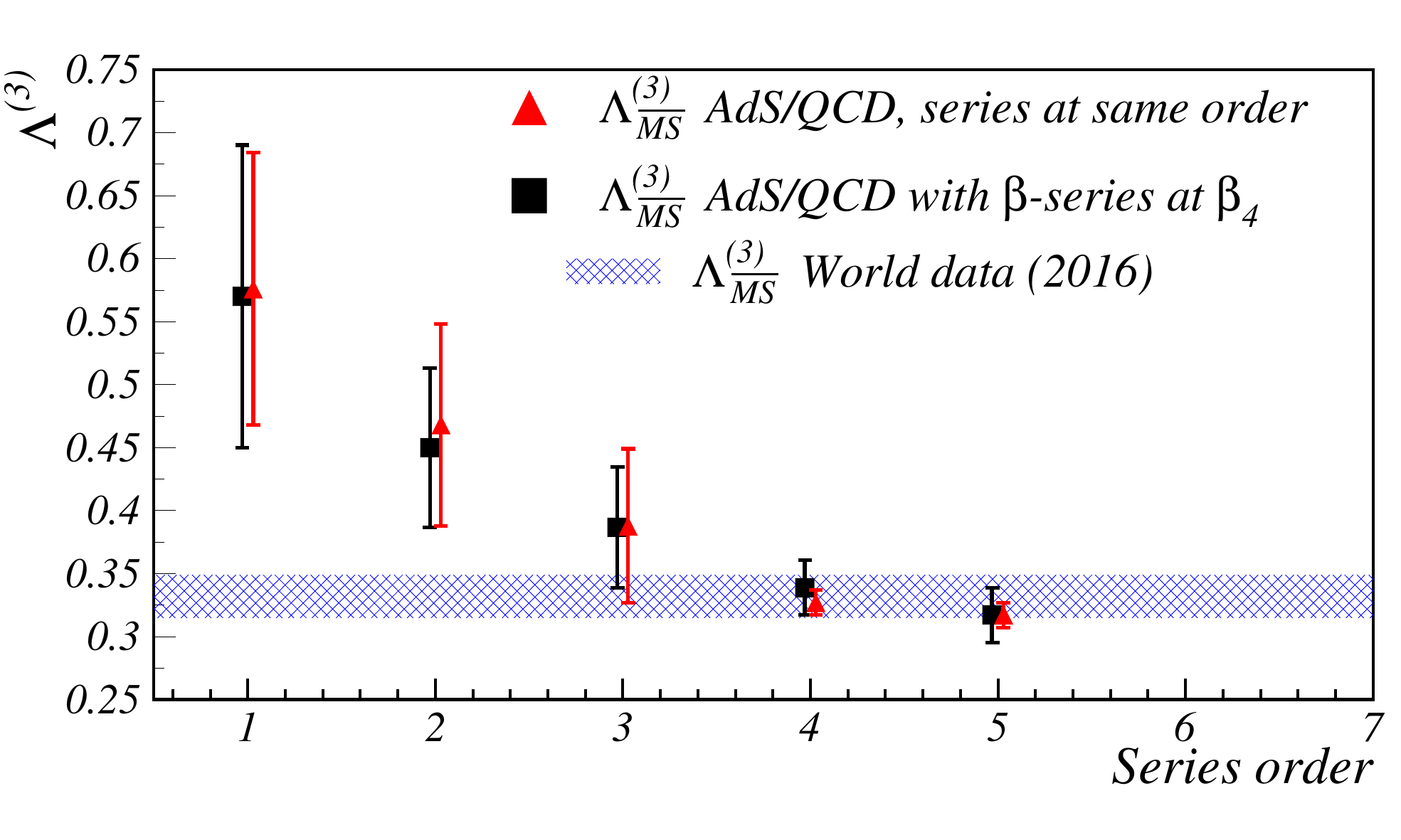}}
\caption{\label{Flo:convergence2} Convergence of our determination of $\Lambda_{\overline{MS}}$ 
 as a function of the Bjorken series order (squares). $\alpha_{\overline{MS}}^{pQCD}$
is computed at order $\beta_{4}$ and for $n_{f}=3$.  
The triangles represent the results when both the $\beta$- and  the Bjorken series are
computed at the same order.  
The error bars include only the uncertainty from the series truncation.
The blue band gives the latest world data.}
\end{figure}

The convergence with respect to the $\beta$-order is shown in Fig. \ref{Flo:convergence_beta} for the Bjorken series calculated
at order $\left(\alpha_{\overline{MS}}^{pQCD}\right)^{4}$. 
This series oscillates but nevertheless converges well. The convergence with respect to the Bjorken
series order is shown in Fig. \ref{Flo:convergence2} for $\alpha_{\overline{MS}}^{pQCD}$
calculated at order $\beta_{4}$. The overall convergence of our method is estimated with both the $\beta$-
and  the Bjorken series calculated at the same order. This is also shown in Fig.  \ref{Flo:convergence2}.
The convergence is slightly faster than the case when the $\beta$-series is kept at  order $\beta_4$.

%
%

%
%

\section{Conclusion}
We have updated the analysis initially reported in Ref.~\cite{Deur:2014qfa} . 
The improved prediction $\Lambda_{\overline{MS}} = 0.339 \pm 0.019$ GeV obtained
from matching  the light-front holographic QCD (LFHQCD)  predictions, constrained by the superconformal 
algebraic structure, and the perturbative QCD  five-loop  computation,  is in excellent
agreement with the value from the present world data, $\Lambda_{\overline{MS}}^{\rm PDG} = 0.332 \pm 0.017$
GeV.  The GeV units conventionally used for mass involves physics external to QCD.   
QCD only predicts dimensionless ratios of masses such as $m_\rho / m_p$.   We thus cast our main result as the ratio: 
 $${\Lambda_{\overline {MS}} / m_p}  = 0.361  \pm 0.020,$$ 
to be compared with the world data ${\Lambda_{\overline {MS}} / m_p} =  0.354  \pm 0.017$.
The uncertainty of our result, which has decreased by almost a factor of 60\% compared to 
Ref.~\cite{Deur:2014qfa}, is now in par with the global world data. 
The improved precision allows one to test more stringently the LFHQCD approach to QCD, 
with a precision  comparable  with that of the most sensitive tests of QCD.

Our method is applicable for setting the perturbative QCD scale $\Lambda_s$ in any
renormalization scheme. We have used the ${\overline{MS}}$ scheme since this has been 
the conventional choice for pQCD analyses.

We performed a convergence analysis that validates and improves the method. 
This could not be done without the newly available 5-loop calculation, as can be seen by removing the last 
point of Fig.~\ref{Flo:convergence2}.
The convergence of the method is satisfactory overall,  for both the $\beta$-series
and the   pQCD  prediction for the Bjorken sum rule.  The largest uncertainty
stems from the truncation of the Bjorken sum pQCD series. A calculation
of its next term --presently only estimated--  and the application of the 
Principle of Maximum Conformality (PMC)~\cite{Mojaza:2012mf,Brodsky:2013vpa} 
would be valuable for  further improving the accuracy of the method discussed here. The uncertainty from the 
determination of  the mass scale $\kappa$ from hadronic spectroscopy contributes similarly.  
Thus a reduction in the uncertainty of its value will provide an even more accurate
holographic prediction for $\Lambda_{\overline{MS}}$. 

The excellent agreement between  the light-front holographic prediction and the world data 
validates with high accuracy the relevance of the gauge/gravity approach to nonperturbative strong interaction 
phenomena  and the constraints imposed by superconformal quantum mechanics.
The LFHQCD approach is a remarkable advance for hadron physics since it provides a direct connection between 
the mass scale $\kappa$, underlying the masses of the proton and other hadrons, with the mass scale 
$\Lambda_{\overline {MS}}$  underlying perturbative QCD.   It  also leads to a description of the QCD running 
coupling at all scales and determines a transition scale between the nonpertubative and perturbative domains.   
These advances have been long-term goals of hadron physics.

 \section*{Acknowledgments}
This material is based upon work supported by the U.S. Department of Energy, Office of Science, Office of Nuclear Physics under contract DE--AC05--06OR23177. This work is also supported by the Department of Energy  contract DE--AC02--76SF00515. SLAC-PUB-16772.

\end{document}